\documentclass[12pt,preprint]{aastex}
\usepackage{natbib}

\begin{document}

\title{A Two-Component Power Law Covering Nearly Four Orders of Magnitude in
the Power Spectrum of Spitzer Far-Infrared Emission from the Large
Magellanic Cloud}

\author{David L. Block}
\affil{School of Computational and Applied Mathematics, University of
the Witwatersrand, Private Bag 3, WITS 2050, South Africa}

\author{Iv\^anio Puerari}
\affil{Instituto Nacional de Astrof\'\i sica, Optica y Electr\'onica,
Calle Luis Enrique Erro 1, 72840 Santa Mar\'\i a Tonantzintla, Puebla,
M\'exico.}

\author{Bruce G. Elmegreen}
\affil{IBM T. J. Watson Research Center, 1101 Kitchawan Road, Yorktown
Heights, New York 10598 USA, bge@us.ibm.com}

\author{Fr\'ed\'eric Bournaud}
\affil{CEA, IRFU, SAp, F-91191 Gif-sur-Yvette, France.}

\begin{abstract}
Power spectra of Large Magellanic Cloud (LMC) emission at 24, 70 and
160 $\mu$m observed with the Spitzer Space Telescope have a
two-component power-law structure with a shallow slope of $-1.6$ at low
wavenumber, $k$, and a steep slope of $-2.9$ at high $k$. The break
occurs at $k^{-1}\sim100-200$ pc, which is interpreted as the
line-of-sight thickness of the LMC disk. The slopes are slightly
steeper for longer wavelengths, suggesting the cooler dust emission is
smoother than the hot emission. The power spectrum covers $\sim3.5$
orders of magnitude and the break in the slope is in the middle of this
range on a logarithmic scale. Large-scale driving from galactic and
extragalactic processes, including disk self-gravity, spiral waves and
bars, presumably cause the low-$k$ structure in what is effectively a
two-dimensional geometry. Small-scale driving from stellar processes
and shocks cause the high-$k$ structure in a 3D geometry. This
transition in dimensionality corresponds to the observed change in
power spectrum slope.  A companion paper models the observed power-law
with a self-gravitating hydrodynamics simulation of a galaxy like the
LMC.
\end{abstract}

\section{Introduction}

The Large Magellanic Cloud (LMC) is the first spiral galaxy in which an
elongated feature of stars was identified. In a remarkable drawing of
the LMC as seen with the naked eye made by Sir John Herschel, a bar is
clearly shown, which Sir John termed an ``axis of light''. Also
faithfully represented in that drawing is a prominent spiral arm to the
north, as well as two ``embryonic'' arms. The LMC belongs to the de
Vaucouleurs classification bin SB(s)m, and is approximately 50 kpc
distant \citep[based on the distance modulus $m-M=18.50 \pm 0.10$,
following][]{freedmanetal01}. Studies of the LMC concur that the
inclination to the line-of-sight of the (warped) disk lies in the range
30$^{\circ}$--50$^{\circ}$ while the position angle of the line of
nodes lies between 120$^{\circ}$ and 150$^{\circ}$; an excellent review
is provided by \citet{vandermareletal08}. The study by \citet{cioni01}
using DENIS and 2MASS surveys yields an inclination of 35$^{\circ}$ and
position angle of 125$^{\circ}$, which are the values adopted in this
study.

The infrared emission from dust in the diffuse interstellar medium of
our Galaxy has been extensively studied from IRAS to the Herschel Space
Observatory. The quantitative Galactic model of \citet{draine} consists
of a mixture of amorphous silicate grains and carbonaceous grains, each
with a wide size distribution: sizes range from molecules containing
tens of atoms to large grains greater than 1 $\mu$m. Dust grain
temperatures at the surfaces of giant molecular clouds in our Galaxy
are hotter than in their cold interiors \citep{greenbergli96}, varying
from $\sim$ 15 K to $\sim$ 7 K, respectively. Cold dust grains are a
crucial component in the ISM of galaxies (e.g., Block et al. 1994).
With regard to the dust temperature - metallicity relations in external
systems, \citet{engelbracht08} find an anticorrelation between dust
temperature and metallicity. Equilibrium dust temperatures are $\sim$
23 K for solar-type metallicities, increasing to 40 K at a low
metallicity of $12+\log(O/H)\sim8$ (see their Figure 5). First insights
from a Herschel study of Messier 33, with its strong metallicity
gradient, show that dust temperatures are $\sim$ 25 K in the central
portion of the stellar disk, declining to $\sim$ 15 K in the outer
domains \citep[Figure 2 in][]{braine10}.

Here we study the power spectrum (PS) of FIR dust emission from the
LMC. Power spectra of emission maps are an important but poorly
understood diagnostic for interstellar structures and the motions that
cause them. PS of HI emission from sections of the Milky Way are
approximately power laws throughout the entire range of observed
spatial frequencies \citep{crov83,green93,dickey01,khalil}. Similar
slopes occur for PS of Milky Way CO emission \citep{stut98}, IRAS and
DIRBE 100 $\mu$m emission \citep{gautier,schlegel} and HI absorption
\citep{des00}. These power laws resemble theoretical expectations from
passive gas motions \citep{goldman} and compressions \citep{laz00} in a
turbulent fluid, so it is natural to interpret them in this way
\citep{es04}. The situation is much more complex for entire galaxies,
however, as galaxies undergo rotation and contain supernovae, spiral
waves, and other physical processes that are not present in localized
clouds. Still, \citet{stan99} found a power law covering 2.5 orders of
magnitude in the PS of HI emission from the Small Magellanic Cloud
(SMC). Subsequent analysis of SMC dust emission \citep{stan00} gave a
similar result. In all of these cases, the two-dimensional PS is
approximately a single power law with a slope between $-2.8$ and $-3$,
depending on the tracer and the depth of the velocity slice
\citep[e.g.,][]{dickey01}.

The PS of HI emission from the Large Magellanic Cloud (LMC) is
significantly different from these others. \citet{elme01} found a
two-component power-law with a break at a wavenumber that is consistent
with the inverse of the disk thickness. They proposed that turbulent
processes and other gas motions are approximately 2D on scales larger
than the disk thickness, and similar processes are 3D on scales smaller
than the disk thickness. PS for integrated emission from turbulent gas
has a slope that differs by 1 in these two cases, with the steeper
slope corresponding to 3D motions at large wavenumbers
\citep{laz00,laz01}. In the LMC, the break scale increases with radius,
indicative of a reasonable level of flare in the HI disk. Subsequent
models of disk PS by \citet{padoan01} confirmed this interpretation.

The same two-part PS was also found for discrete clouds in the Milky
Way. \citet{miv03} found it in HI emission from the Ursa Major Galactic
cirrus and modeled it with a fractal Brownian motion cloud. Their model
suggested that the break in the PS occurs at a wavenumber $k\sim1/(2D)$
for cloud depth $D$, and that the power-law slope changes by 1 at the
break. \citet{ingalls04} observed a two-component PS using Spitzer IR
observations of a molecular cloud in the GUM nebula. They found that
the PS slope changed from $-2.6$ at wavenumbers less than
$2\times10^{-3}$ arcsec$^{-1}$ to $-3.5$ at wavenumbers greater than
$4\times10^{-3}$ arcsec$^{-1}$. The inferred cloud depth was 0.3 pc.

A break in the PS of HI emission has not generally been observed in
galaxies other than the LMC. \citet{dut08} studied HI emission from the
face-on galaxy NGC 628 and derived a PS slope of $-1.7$ without a
break. Their PS covered only a factor of 10 in scale, which may be too
small to resolve out two separate components. \citet{dut09a} saw a
break in NGC 1058 with a steepening of the slope from $-1$ to $-2.5$ at
an inverse wavenumber of 1.5 kpc, which they converted to a disk
thickness of 490 pc. Their PS also spanned a factor of 10 in scale, but
the galaxy may be too distant (10 Mpc) to resolve the disk thickness
with HI (their angular resolution was $13.6^{\prime\prime}$,
corresponding to 660 pc). PS of H$\alpha$ and HI emission from several
dwarf galaxies showed power laws and no breaks
\citep{willett05,begum,dut09b}. In the SMC there is no break either.
For the SMC, this lack of a break could be because the line-of-sight
depth is comparable to the transverse length \citep{westerlund97}. This
could be the case for the dwarf galaxies also \citep{roy}.

PS miss the break if the high frequency part is dominated by the point
spread function of bright unresolved emission sources. These sources
introduce a Gaussian-shaped dip at high frequency in the PS
\citep[e.g.,][]{dickey01}. Models of this dip are shown in
\citet{leitner2} and \citet{block09}. To see a double power law, the
part at high frequency has to extend for a long wavenumber range from
the inverse thickness of the disk to the inverse spatial scale of the
telescope resolution. If the thickness is smaller than the scale of the
resolution, then the high frequency power law from 3D turbulence cannot
be observed. This problem is present for conventional
$\sim10^{\prime\prime}$ HI resolution in galaxies at distances greater
than 2 Mpc. Our Spitzer infrared observations of the LMC have a
resolution scale at least 10 times smaller than the disk thickness, so
we should be able to see any high-frequency power law clearly.

Optical observations of galaxy PS have sufficient resolution to see the
thickness, but unless there are large numbers of stellar groupings with
sizes spanning a wide range around the disk thickness, the clarity of
the PS break will be poor.  A tentative PS break was observed optically
in NGC 5055 \citep{leitner}, but not in NGC 628 \citep{regan}. A break
in the slope of the autocorrelation function for young stars in several
nearby galaxies was inferred to mark a change in geometric properties
of the gas by \citet{odekon08}.

Spitzer Space Telescope observations have such high angular resolution
that 3.5-4 orders of magnitude in spatial scale can be covered for the
LMC. This is much larger than is possible for any other galaxy in any
other gas tracer at the present time. The smallest resolvable scale is
only a few parsecs, which is much smaller than the disk thickness, so
there is no contamination in the PS from the point spread function.
Thus the FIR observations of the LMC are ideal for an unprecedented
view of correlated structures in a galaxy. In a companion paper
\citep{bour10}, we describe self-gravitating hydrodynamic simulations
of galaxies like the LMC, with and without energy input from star
formation. We obtain a two component PS for projected density structure
in the simulations too, and we examine the velocities and energy
sources that lead to this structure.

In what follows, the observations are presented in Section \ref{data},
the PS are shown in Section \ref{ps}, while Section \ref{disc} contains
a discussion of the implications of our results. The conclusions are in
Section \ref{conc}.

\section{Observations}
\subsection{Data}\label{data}

Our data for the LMC come from SAGE \citep[``Surveying the Agents of a
Galaxy's Evolution'';][]{meixner}. The images at 24, 70 and 160 $\mu$m
were obtained with MIPS on the Spitzer Space Telescope; they cover
$\sim 8 \times 8$ degrees. The total integration time is 217 hours,
which improves the point source sensitivity over previous surveys
(e.g., IRAS) by 3 orders of magnitude. In Spitzer, all bands sample the
point-spread function at better than the Nyquist frequency (0.4
$\lambda/D$ at 24 and 160 $\mu$m, 0.3 $\lambda/D$ at 70 $\mu$m). Thus,
the instrument allows the telescope-limited resolution of 6$''$,
18$''$, and 40$''$ at 24, 70, and 160 $\mu$m, respectively
\citep{riekeetal04}. The first of these, $6''$, corresponds to a
spatial scale of 1.45 pc in the LMC.  The theoretical Nyquist limit is
2 pixels, but in reality, the true Nyquist limit is 2 resolution
elements.

At 24 $\mu$m and 70 $\mu$m, the images sizes we use are 8192 $\times$
8192 pixels at a scale of $4\farcs$98 and $4\farcs8$ per pixel,
respectively (the SAGE 24 $\mu$m image was binned $2\times2$ to limit
computer memory requirements). At 160 $\mu$m, the image size is 2048
$\times 2048$ pixels at a scale of $15\farcs6/$px. As far as
deprojecting images of the LMC are concerned, there is no unique
``center'' - the dynamical center of the HI is offset by almost a full
degree from the photometric center of the bar \citep{westerlund97}.  We
choose to conduct our deprojections about the dynamical centre of the
LMC, which has right ascension $\alpha = 5^{h} 27.6^{m}$ and
declination $\delta = -69^{\circ}52^{'}$ (J2000.0), following section 7
in \citet{vandermareletal02}.

\subsection{Power Spectra}\label{ps}

\begin{figure*}
    \vspace{15cm}
     \includegraphics{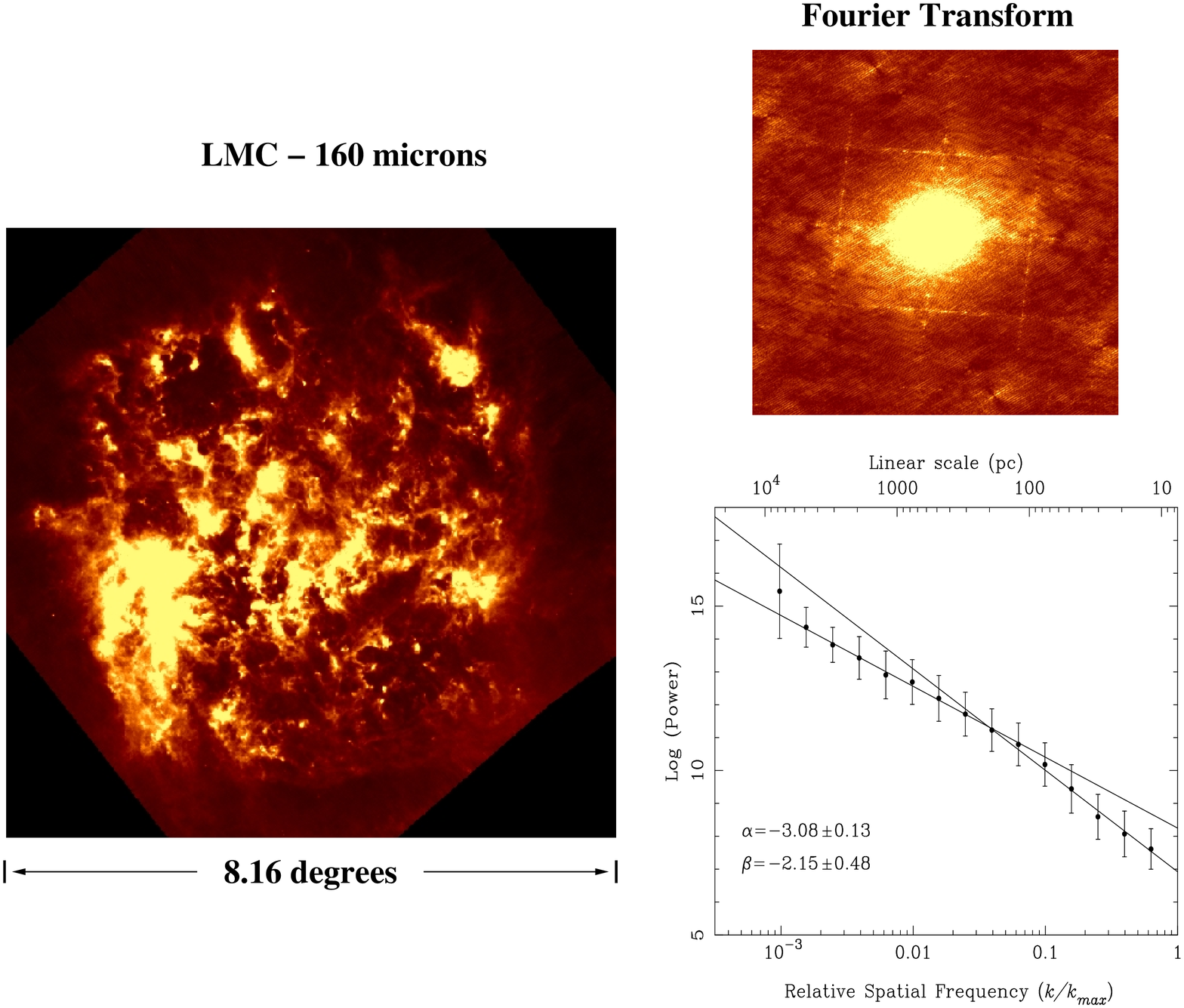}
\caption{Left: The LMC at 160 $\mu$m (North is up, East is left). The
brightest part of this image has an intensity of $\sim1500$ MJy
sr$^{-1}$. The bright cloud in the lower left is the giant molecular
and atomic cloud south of 30 Doradus. Upper Right: Two-dimensional
Fourier transform of the deprojected image. The rectangle betrays low
level striping in the image on the left. Lower Right: 2D PS with power
law fits $\propto k^{\alpha}$ and $\propto k^{\beta}$ at high and low
$k$, respectively; the slopes are indicated at the bottom left. Dots
with error bars ($\pm 1 \sigma$) are averages over a range in $\log k$.
The units of the PS are $({\rm MJy\, sr^{-1}})^2$ for Spitzer
calibrated images. The smallest scale in the PS is 7.6 pc, which is 2
pixels ($31.2''$) and the theoretical Nyquist limit. The PS break at
$\sim200$ pc (top scale) is interpreted as the line-of-sight depth of
the dust disk in the LMC. } \label{powerspectra160}
\end{figure*}

\begin{figure*}
    \vspace{15cm}
     \includegraphics{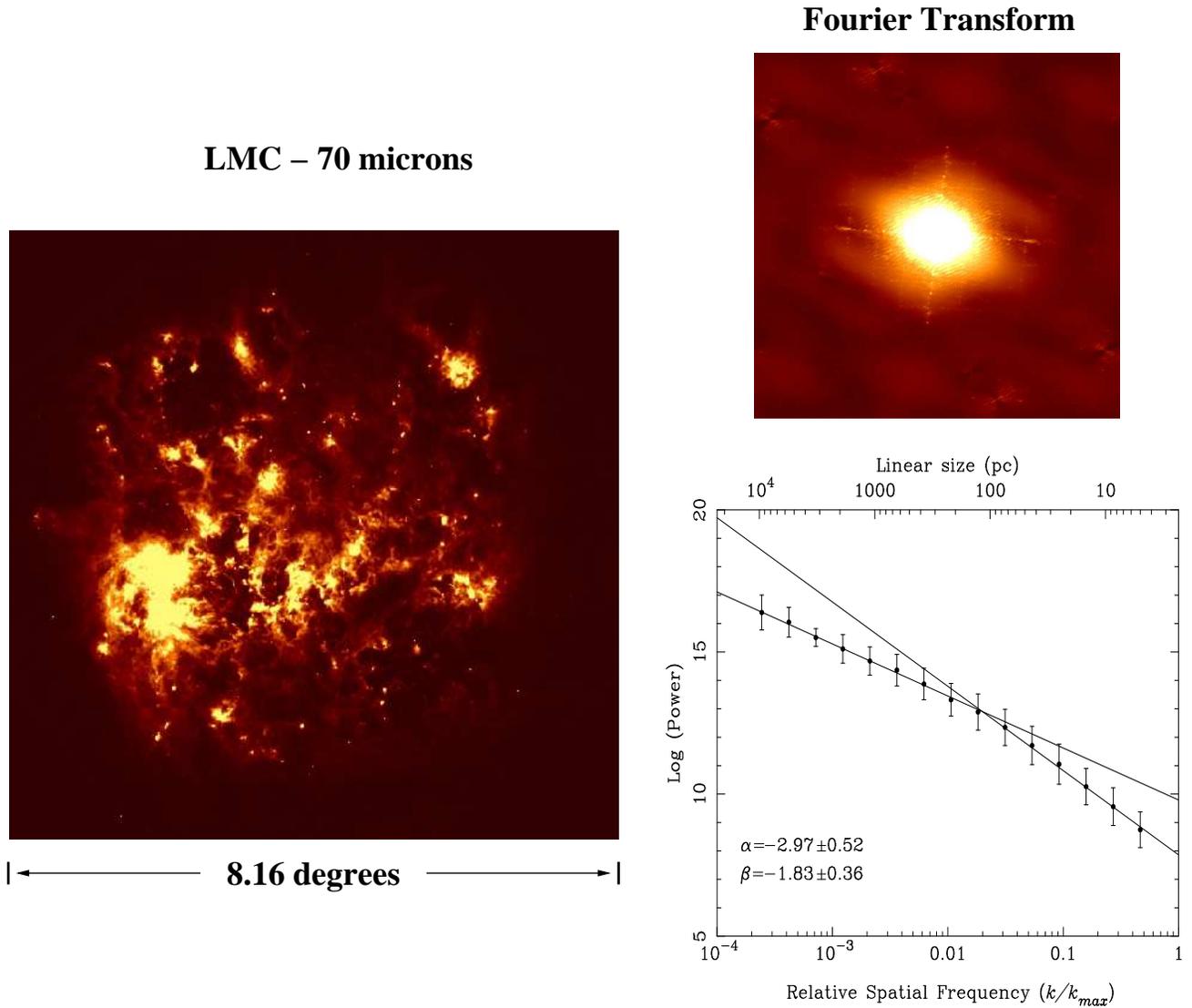}
\caption{Same as Figure 1, but at 70 $\mu$m. The brightest region has
an intensity of 4000 MJy sr$^{-1}$. The smallest scale in the PS is 2.3
pc, which is 2 pixels ($9.6''$) and the theoretical Nyquist limit. The
power spectrum spans 3.5 orders of magnitude.} \label{powerspectra070}
\end{figure*}

\begin{figure*}
    \vspace{15cm}
     \includegraphics{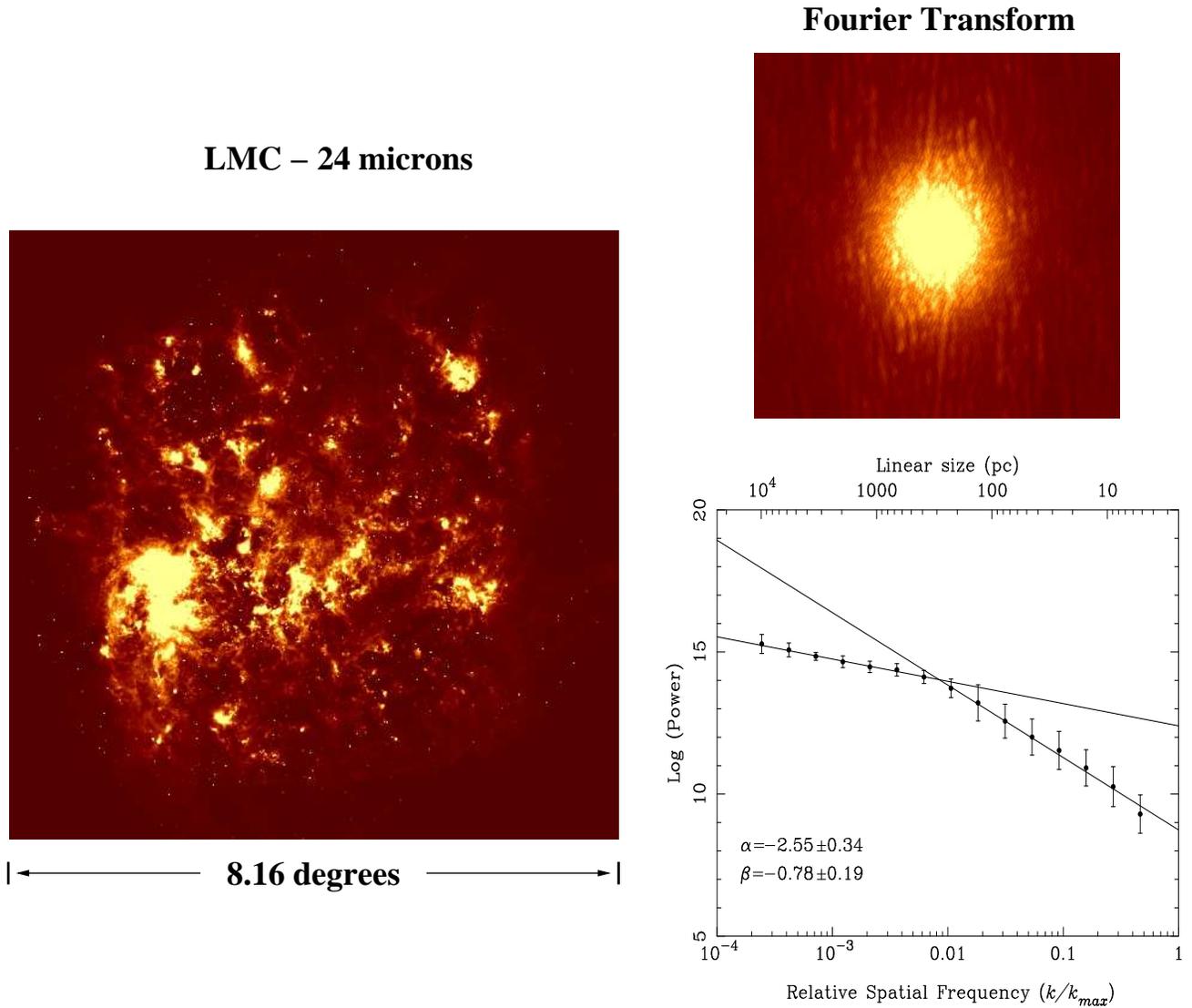}
\caption{Image and PS of the LMC at 24 $\mu$m. The brightest region has
an intensity of 2800 MJy sr$^{-1}$. The smallest scale in the PS is 2.4
pc, which is 2 pixels ($10.0''$) and the theoretical Nyquist limit.}
\label{powerspectra024}
\end{figure*}

Figure \ref{powerspectra160} shows the MIPS image at 160 $\mu$m and the
2D PS at the bottom right. The PS is the sum of the squares of the real
and imaginary parts of the 2D Fourier transform (made with {\tt fourn}
from {\it Numerical Recipes}), averaged over each 2D wavenumber
$k=(k_x^2+k_y^2)^{1/2}$. We divided the PS into 15 wavenumber intervals
and plotted the arithmetic mean for each interval as a dot in Figure
\ref{powerspectra160}. The PS has 2 distinct parts with different power
law slopes. The steep slope at high $k$ is $-3.08\pm0.13$ and the
shallow slope at low $k$ is $-2.15\pm0.48$. The break is at
approximately $1/k=200$ parsecs. Double power laws also occur in the PS
of the other MIPS channels (Figs. \ref{powerspectra070} and
\ref{powerspectra024}). The break in the middle of the PS on a log
scale allows both power laws to be seen clearly; there are $1.5-2$
orders of magnitude of wavenumber on each side for the shorter
wavelength bands.

Emission at 160 $\mu$m arises from cold dust in dark clouds. At 70
$\mu$m, the emission comes from both warm and cold dust, and at 24
$\mu$m the emission is from warm dust and PAH's. The figures indicate
that all of this dust has correlated structure spanning the entire
resolvable disk of the LMC. The PS break is at $100-200$ pc for all 3
MIPS bands, suggesting that the disk thickness is about this value for
each dust component. The least-square slopes are similar but with a
slight progression. At high spatial frequency, they are $-3.08\pm0.13$,
$-2.97\pm0.52$ and $-2.55\pm0.34$ for 160 $\mu$m, 70 $\mu$m and 24
$\mu$m, respectively, and at low spatial frequency, they are
$-2.15\pm0.48$, $-1.83\pm0.36$ and $-0.78\pm0.19$ for these three
passbands. The steepening of the slope with FIR passband indicates that
there is relatively more spectral power on large scales for longer
wavelengths. This suggests that cool dust is more diffuse and less
structured on small scales than warm dust.  This conclusion is true
regardless of the total fluxes and column densities of cool and warm
dust, as these quantities contribute only to the $k=0$ component of the
PS.

For comparison, the 2D PS of HI emission from the LMC has a slope of
$\sim-3.7$ at high spatial frequency and a slope of $\sim-2.7$ at low
spatial frequency \citep{elme01}. The break is at a scale of
$1/k\sim100$ pc, increasing slightly with galactocentric radius. These
slopes are both higher than the corresponding slopes for the dust
emission. The HI PS is most like the 160 $\mu$m PS, although even
steeper, suggesting that HI is even more dominated by large scale
structure than cool dust.

In order to test the robustness of our analysis, we also deprojected
the MIPS images about the bar center as opposed to the dynamical center
of the LMC; the resulting PS at 24 $\mu$m remained unchanged, while the
slopes of the two power laws at 70 $\mu$m and 160 $\mu$m were only
marginally affected. We conclude that choice of center for deprojection
purposes is not responsible for the presence of two distinct power
laws.

\section{Discussion}\label{disc}

The figures indicate that the relative thickness of the LMC dust disk
compared to its diameter is only a few percent.  This value comes from
the ratio of the smallest wavenumber in the PS (the inverse of the
largest scale) to the wavenumber at the break in the PS (the inverse of
the disk thickness). The smallness of this ratio indicates that the LMC
geometry is effectively 2D for some processes.

Turbulence in a strictly 2D medium (infinitely thin) differs from
turbulence in a 3D medium. For incompressible turbulence, an infinitely
thin layer can have an inverse cascade of energy, i.e., energy moves
from smaller to larger scales, and a direct cascade of mean-squared
vorticity, i.e., from larger to smaller scales
\citep{kraichnan,leith,batchelor}. For 3D turbulence, the energy
generally cascades to smaller scales. Thus 2D turbulence on large
scales can be fed by small scale motions, while 3D turbulence on large
scales can only be fed by larger scale motions (in the absence of a
magnetic field or other physical processes that force a connection
between large and small scales).

It would be interesting if an inverse cascade of turbulence occurred in
a very thin galaxy. Then localized energy from gravitational
instabilities on the scale of the Jeans length, which is the disk scale
height, and from supernovae, superbubbles, and so on, could power large
scale motions and density correlations. A galactic disk is not strictly
2D even if it is thin, but the gas motions can still be highly
anisotropic, making the disk effectively 2D. For example, density-wave
and bar-driven streaming motions are often 5 to 10 times faster than
the perpendicular motions that produce the disk thickness.  Large scale
motions may be generated by disk self-gravity, tidal forces from
companion galaxies \citep{mas09}, intergalactic ram pressure
\citep{tonn,dut10}, and other forcings. They contribute to large-scale
structure and to the low-$k$ part of the density PS. More localized
energy input from stars and OB associations should drive 3D motions and
structures that contribute to the high-$k$ part of the density PS
\citep[see also][]{hodge}. Turbulence driven on large scales would have
a shallow PS because of the 2D nature of the resulting flows, while
turbulence driven on small scales would have a steep PS because of the
3D nature of its associated flows.

\citet{bour10} ran simulations of galaxies like the LMC and found a
two-component PS similar to what we observe here. This result occurred
whether or not there was input from supernova energy, suggesting that
gravitational energy alone can drive galactic turbulence over a wide
range of scales. The study also suggested that there may be some
contribution to the large-scale motions from an inverse cascade of
energy in 2D. For example, they found a $k^{-1}$ PS for the
mean-squared vorticity, as in numerical experiments of strictly 2D
turbulence in protoplanetary disks \citep{peterson}. Thus the LMC PS
could be the result of gravity-driven turbulence and other large-scale
motions contributing to gas structure in an effectively 2D medium wider
than the disk thickness, supernovae, cloud-scale gravity and other
small-scale motions contributing to turbulence in a 3D medium within
the disk thickness, and an inverse cascade of energy at the thickness
scale up to larger scales, contributing to turbulence throughout the
disk.

\section{Conclusions}\label{conc}

The LMC is the only disk galaxy where we can currently study
interstellar motions on a range of scales covering a factor of $10^4$
or more. An essential component of this study is the Spitzer Space
Telescope, which provides the highest possible angular resolution for
large-scale maps of interstellar emission.  Using the SAGE images at 24
$\mu$m, 70 $\mu$m, and 160 $\mu$m, we derived 2D power spectra of dust
emission over the whole LMC disk. The results showed one power law
covering a factor of 100 in scales at low spatial frequency and
another, steeper, power law covering another factor of 100 in scales at
high frequency. The low frequency power law had been observed before
with HI emission, but the high frequency power law was more limited
before, spanning only a factor of $\sim10$ in range \citep{elme01}. The
current observations confirm the two-component nature of the
interstellar motions, suggesting that gas motions and turbulence are
two-dimensional on scales larger than the disk thickness, and
three-dimensional on scales smaller than the disk thickness. The
line-of-sight thickness is measured at the break point, and is between
100 pc and 200 pc, depending on the wavelength of observation.

\noindent{\bf Acknowledgments} We are grateful to G.G. Fazio and J.
Hora for providing us with timely access to the SAGE Survey images of
the LMC after their public release. We are indebted to S. Stanimirovic,
L. Staveley-Smith and S. Kim for sending us their HI images of the SMC
and LMC for testing our PS code. We are grateful to J. Scalo for
comments on an early version of this manuscript. A note of deep
appreciation is expressed by DLB to AVENG and to Mr. F. Titi for their
sponsorship of his research. DLB is indebted to Roger Jardine, Kim
Heller and to the AVENG Board of Trustees. This research is partially
supported by the Mexican Foundation CONACYT.  This study is based on
observations made with the Spitzer Space Telescope, which is operated
by the Jet Propulsion Laboratory, California Institute of Technology
under a contract with NASA.

\vfill\eject
\null

\end{document}